\documentclass[preprint,superscriptaddress,amsmath,amssymb,aps,notitlepage,floatfix,longbibliography]{revtex4-1}
\usepackage{graphicx}
\usepackage{color}

\usepackage{dcolumn}
\usepackage{upgreek}
\usepackage{gensymb}
\usepackage{float}
\usepackage{bm}
\usepackage{xcolor}
\usepackage{physics}

\usepackage{xr}
\makeatletter
\newcommand*{\addFileDependency}[1]{
  \typeout{(#1)}
  \@addtofilelist{#1}
  \IfFileExists{#1}{}{\typeout{No file #1.}}
}
\makeatother

\newcommand*{\myexternaldocument}[1]{
    \externaldocument{#1}
    \addFileDependency{#1.tex}
    \addFileDependency{#1.aux}
}

\myexternaldocument{supps_072020}

\begin{document}

\title{Acoustic spin Hall effect in strong spin-orbit metals}

\author{Takuya Kawada}
\affiliation{Department of Physics, The University of Tokyo, Tokyo 113-0033, Japan}

\author{Masashi Kawaguchi}
\email[]{masashi.kawaguchi@phys.s.u-tokyo.ac.jp}
\affiliation{Department of Physics, The University of Tokyo, Tokyo 113-0033, Japan}

\author{Takumi Funato}
\affiliation{Department of Physics, Nagoya University, Nagoya 464-8602, Japan}

\author{Hiroshi Kohno}
\affiliation{Department of Physics, Nagoya University, Nagoya 464-8602, Japan}

\author{Masamitsu Hayashi}
\email[]{hayashi@phys.s.u-tokyo.ac.jp}
\affiliation{Department of Physics, The University of Tokyo, Tokyo 113-0033, Japan}

\newif\iffigure
\figurefalse
\figuretrue

\date{\today}

\maketitle

\textbf{
We report on the observation of the acoustic spin Hall effect that facilitates lattice motion induced spin current via spin orbit interaction (SOI).
Under excitation of surface acoustic wave (SAW), we find a spin current flows orthogonal to the propagation direction of a surface acoustic wave (SAW) in non-magnetic metals.
The acoustic spin Hall effect manifests itself in a field-dependent acoustic voltage in non-magnetic metal (NM)/ferromagnetic metal (FM) bilayers.
The acoustic voltage takes a maximum when the NM layer thickness is close to its spin diffusion length, vanishes for NM layers with weak SOI and increases linearly with the SAW frequency.
To account for these results, we find the spin current must scale with the SOI and the time derivative of the lattice displacement.
Such form of spin current can be derived from a Berry electric field associated with time varying Berry curvature and/or an unconventional spin-lattice interaction mediated by SOI.
These results, which imply the strong coupling of electron spins with rotating lattices via the SOI, show the potential of lattice dynamics to supply spin current in strong spin orbit metals.
}

\clearpage

Spin current represents a flow of spin angular momentum carried by electrons.
The spin Hall effect\cite{dyakonov1971jetp} allows electrical generation of spin current in materials with strong spin orbit interaction (SOI)\cite{galitski2013nature}. 
The spin Hall angle, a material parameter that characterizes charge to spin conversion efficiency, scales with the longitudinal resistivity and the spin Hall conductivity\cite{hoffmann2013ieee}.
For the intrinsic spin Hall effect, the spin Hall conductivity is determined by the electron band structure\cite{sinova2004prl,guo2008prl} (\textit{i.e.}, the Berry curvature of the bands near the Fermi level) and the SOI of the host material.
As spin current can be used to control the direction of magnetization of a ferromagnetic layer placed adjacent to the spin source, developing materials and means to create it with high efficiency are the forefront of modern Spintronics\cite{miron2011nature,liu2012science,manchon2019rmp}.


Recent studies have shown that not only electrons but other degrees of freedom can generate spin current.
Precessing magnetization pumps out spin current from magnetic materials, a mechanism known as spin pumping\cite{mizukami2002prb,tserkovnyak2002prl,saitoh2006apl}.  
In the spin Seebeck effect\cite{uchida2010nmat,bauer2012nmat}, a temperature gradient applied to a magnetic material induces a magnon population gradient and the associated diffusion spin current.
Spin current can also be produced from exchange of angular momentum between a rotating body and electrons, an effect referred to as spin-rotation coupling\cite{matsuo2013prb}.
The effect has been observed in liquid metals\cite{takahashi2016nphys} and non-magnetic light metals (\textit{e.g.}, Cu)\cite{kobayashi2017prl}.
Generation of spin current via spin pumping, spin Seebeck effect and spin-rotation coupling do not require large SOI of the host material.

Here we show a profoundly different approach to generate spin current.
We find a spin current directly emerges from the dynamics of lattice via SOI.
Similar to the spin Hall effect where a spin current flows transverse to electrical current, a spin current develops orthogonal to the propagation direction of a surface acoustic wave (SAW) in non-magnetic metals.
The efficiency to generate spin current is proportional to the spin Hall angle and may be influenced by a factor that depends on the film structure.
To account for the experimental results, we find the spin current must scale with the SOI and the time derivative of the lattice displacement.

Thin film heterostructures are grown on piezoelectric LiNbO$_3$ substrates using radio frequency (rf) magnetron sputtering.
The film structure is sub./X($d$)/CoFeB(1)/MgO(2)/Ta(1) with X=W, Pt, Ta and Cu (thickness in unit of nanometers). 
The heterostructures are referred to as X/CoFeB bilayers hereafter. 
Standard optical lithography is used to pattern Hall bars from the film and electrodes/interdigital transducers (IDTs)\cite{white1965apl} from conducting metals (see Methods for the details of sample preparation).

\begin{figure}[t]
	\includegraphics[scale=1.0]{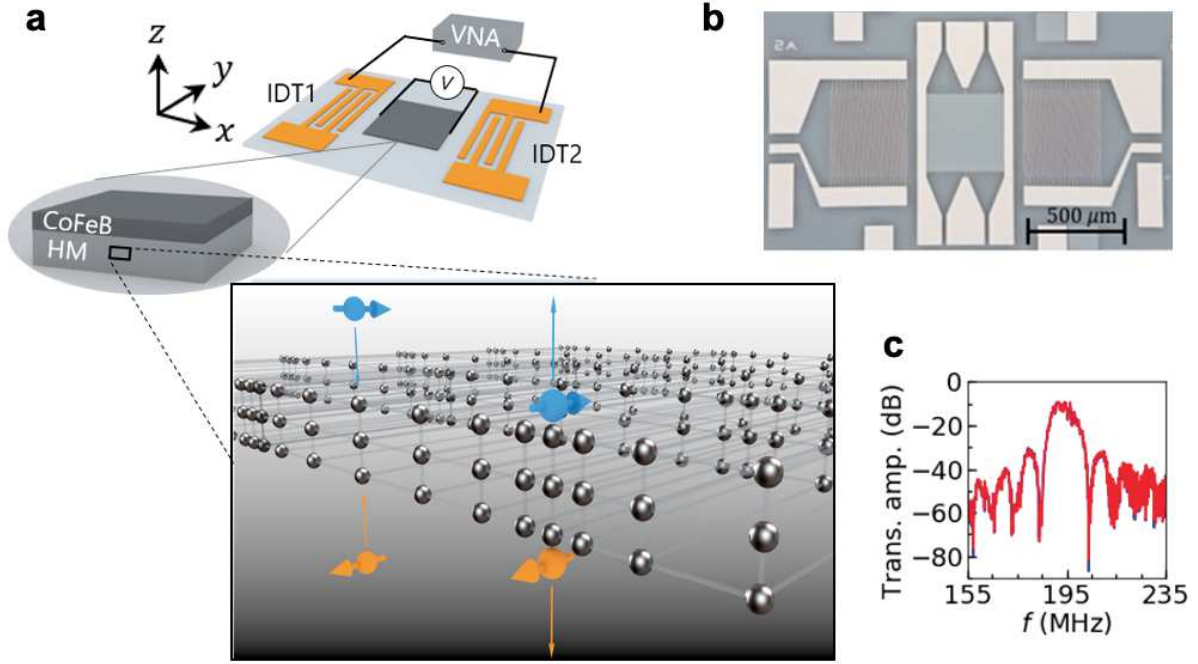}
	\caption{\textbf{Experimental setup to probe the acoustic spin Hall effect.} (a) Schematic illustration of the experimental setup including the substrate, the film, the IDTs and the VNA. The bottom image illustrates lattice motion induced spin current, \textit{i.e.}, the acoustic spin Hall effect. Spin current flows orthogonal to the SAW propagation. (b) Representative optical microscopy image of the device. The bright regions are the electrodes and the gray square at the center is the Hall bar made of the film. (c) SAW transmission amplitude from IDT1 to IDT2 (IDT2 to IDT1) is plotted as a function of frequency ($f$) by the blue (red) line. A Hall bar made of W(2.4)/CoFeB bilayer is placed between the IDTs. 
	\label{fig:schematics}}
\end{figure}

The experimental setup and the coordinate system are schematically illustrated in Fig.~\ref{fig:schematics}(a). 
The Hall bar is placed between the two IDTs. 
Figure~\ref{fig:schematics}(b) shows a representative optical microscope image of the device. 
A vector network analyzer (VNA) is connected to the IDTs to excite a Rayleigh-type SAW from one end and to detect its transmission at the other end.
Figure~\ref{fig:schematics}(c) shows typical transmission spectra with a W/CoFeB bilayer placed between the IDTs.
The transmission amplitude takes a maximum at $\sim$194 MHz, which corresponds to the fundamental excitation frequency of the SAW ($f_\mathrm{SAW}$) defined by the geometry of the IDTs and the sound velocity of the substrate.

The acoustoelectric properties of the films are studied as a function of magnetic field.
A continuous rf signal with fixed frequency $f$ and power $P$ is fed from one of the VNA ports to the corresponding IDT, which launches a SAW along $x$ that propagates to the film and induces lattice motion. 
The longitudinal (along $x$) and transverse (along $y$) voltages of the Hall bar, defined as $V_{xx}$ and $V_{yx}$, respectively, are measured during the SAW excitation.
Since $V_{xx}$ and $V_{yx}$ contain similar information, here we focus on the results of $V_{xx}$; see supplementary material section~\ref{sec:supp:results} for the characteristics of $V_{yx}$. 
In order to extract the voltage originating from the SAW, we subtract the average voltage measured under off-resonance conditions ($f \neq f_\mathrm{SAW}$) and obtain the acoustic voltage $\Delta V_{xx} \equiv V_{xx} - \langle {V}_{xx}^\mathrm{off} \rangle$.
$\langle {V}_{xx}^\mathrm{off} \rangle$ is the average value of $V_{xx}$ when $f$ is set far from $f_\mathrm{SAW}$ (see Methods for the details).
We apply an in-plane magnetic field of magnitude $H$ during the voltage measurements. 
The angle between the field and the $x$-axis is defined as $\varphi_H$.
As the magnetic easy axis of the CoFeB layer points along the film plane and the in-plane magnetic anisotropy is weak, we assume the orientation of the magnetization follows that of the magnetic field.   


\begin{figure}[b]
	\includegraphics[scale=1.0]{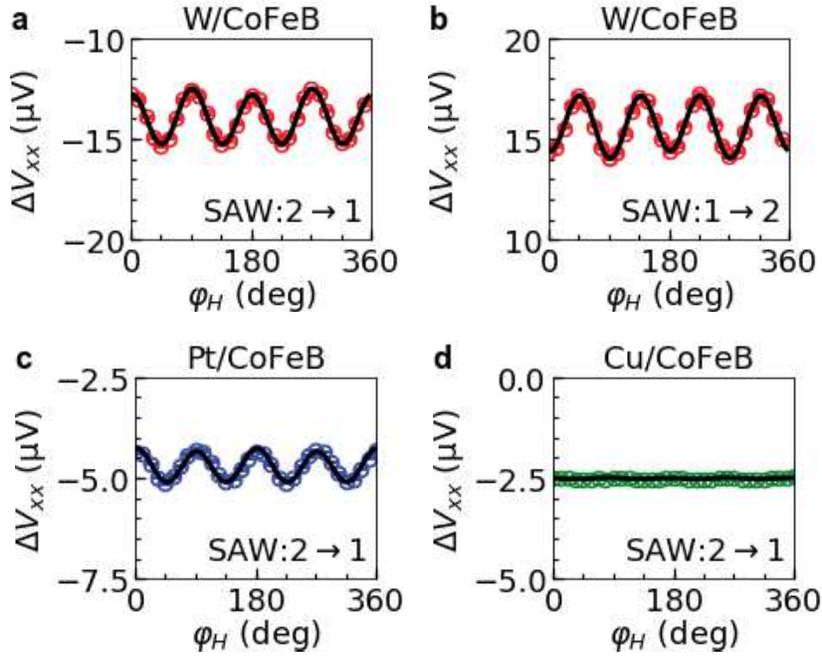}
	\caption{\textbf{Field angle dependence of the acoustic voltage.} (a-d) Magnetic field angle ($\varphi_H$) dependence of $\Delta V_{xx}$ when a rf signal of $f\sim f_\mathrm{SAW}$ and $P\sim$10 dBm is applied to IDT2 (a,c,d) and IDT1 (b). 
	Films placed between the IDTs are W(1.8)/CoFeB (a,b), Pt(2.0)/CoFeB (c), and Cu(1.8)/CoFeB (d) bilayers. The error bars represent standard deviation of the repeated measurements. The black lines show fit to the data with Eq.~(\ref{eq:sawphe}). 
	\label{fig:sawvol}}
\end{figure}

Figures~\ref{fig:sawvol}(a,c,d) show the field angle ($\varphi_H$) dependence of $\Delta V_{xx}$ for W/CoFeB, Pt/CoFeB and Cu/CoFeB bilayers when a rf signal of $f\sim f_\mathrm{SAW}$ and $P\sim$10 dBm is applied to IDT2.
For W/CoFeB and Pt/CoFeB bilayers, $\Delta V_{xx}$ shows a sinusoidal variation with a period of 90$^\circ$.
Note that the sign (\textit{i.e.}, the phase) of the sinusoidal variation is the same for the two bilayers although the sign of the spin Hall angle is opposite between Pt and W\cite{hoffmann2013ieee}.
In contrast, no such variation is found for the Cu/CoFeB bilayer.
Figure~\ref{fig:sawvol}(b) shows $\Delta V_{xx}$ vs. $\varphi_H$ of the W/CoFeB bilayer when the rf signal is applied to IDT1.
Clearly, the mean offset voltage and the sinusoidal variation change their signs as the SAW propagation direction is reversed.
Similar features are observed for the Pt/CoFeB bilayers.

We fit the $\varphi_H$ dependence of $\Delta V_{xx}$ with the following function:
\begin{equation}
\begin{aligned}
\Delta V_{xx} &= \Delta V_{xx}^0 +  \Delta V_{xx}^{2\varphi} \cos^2 \varphi_H + \Delta V_{xx}^{4\varphi} \sin^2 2\varphi_H,
\label{eq:sawphe}
\end{aligned}
\end{equation}
where $\Delta V_{xx}^{n\varphi}$ ($n=$2,4) represents the coefficient of the sinusoidal function with a period of $360^\circ / n$ and $\Delta V_{xx}^{0}$ is the $\varphi_H$-independent component.
$\Delta V_{xx}^{0}$ is proportional to what is known as the acoustic current, which originates from rectification of the localized electric field and charge density\cite{weinreich1957pr}.

\begin{figure}[t]
	\includegraphics[scale=1.0]{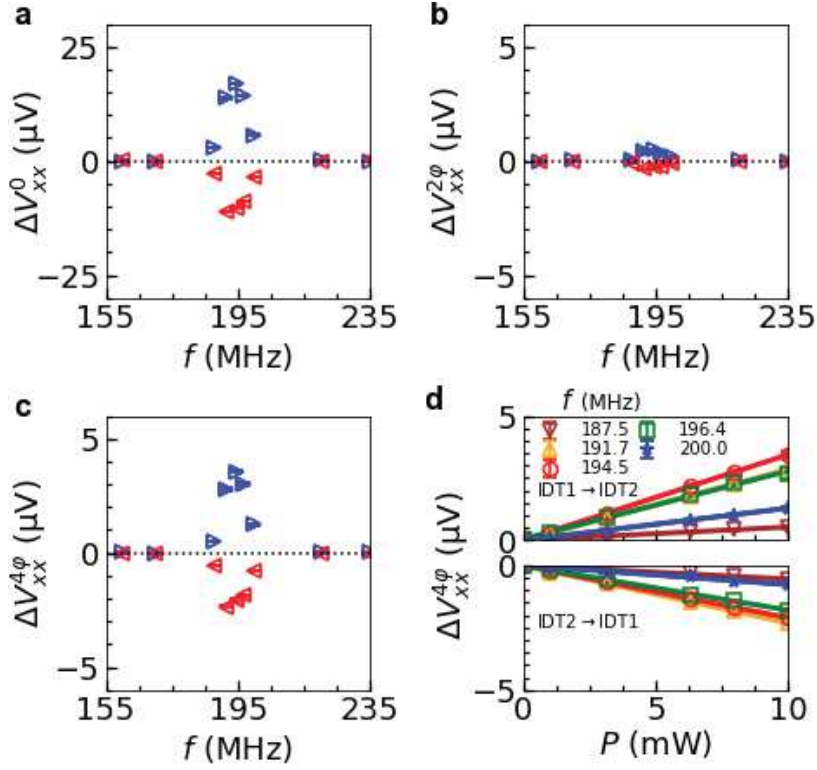}
	\caption{\textbf{Resonant excitation of the acoustic voltage.} (a-c) RF frequency ($f$) dependence of $\Delta V_{xx}^{0}$ (a), $\Delta V_{xx}^{2\varphi}$ (b) and $\Delta V_{xx}^{4\varphi}$ (c). The blue (red) triangles represent results when the rf signal is applied to IDT1 (IDT2). The rf power ($P$) is fixed to $\sim10$ dBm. (d) $P$ dependence of $\Delta V_{xx}^{4\varphi}$ when $f$ is varied. The solid lines show fit to the data with a linear function. Upper and lower panels show results when a rf signal is applied to IDT1 and IDT2, respectively. (a-d) The error bars show fitting errors of $\Delta V_{xx}$ with Eq.~(\ref{eq:sawphe}). Data presented are obtained using W(2.4)/CoFeB bilayer. 
	\label{fig:wcfbfreq}}
\end{figure}

The $f$ dependence of $\Delta V_{xx}^{0}$ is plotted in Fig.~\ref{fig:wcfbfreq}(a). 
$\Delta V_{xx}^{0}$ takes a peak at $f\sim$194 MHz, which corresponds to $f_\mathrm{SAW}$ (see Fig.~\ref{fig:schematics}(c)), and changes its sign as the SAW propagation direction is reversed\cite{kawada2019prb}.
The $f$ dependence of $\Delta V_{xx}^{2\varphi}$ and $\Delta V_{xx}^{4\varphi}$ are shown in Figs.~\ref{fig:wcfbfreq}(b) and \ref{fig:wcfbfreq}(c), respectively.
$\Delta V_{xx}^{4\varphi}$ is significantly larger than $\Delta V_{xx}^{2\varphi}$ and shows a clear peak at $f \sim f_\mathrm{SAW}$, suggesting that its appearance is associated with the excitation of SAW.
The rf power ($P$) dependence of $\Delta V_{xx}^{4\varphi}$ is shown in Fig.~\ref{fig:wcfbfreq}(d). 
$\Delta V_{xx}^{4\varphi}$ increases linearly with $P$.

\begin{figure}[t]
	\includegraphics[scale=1.0]{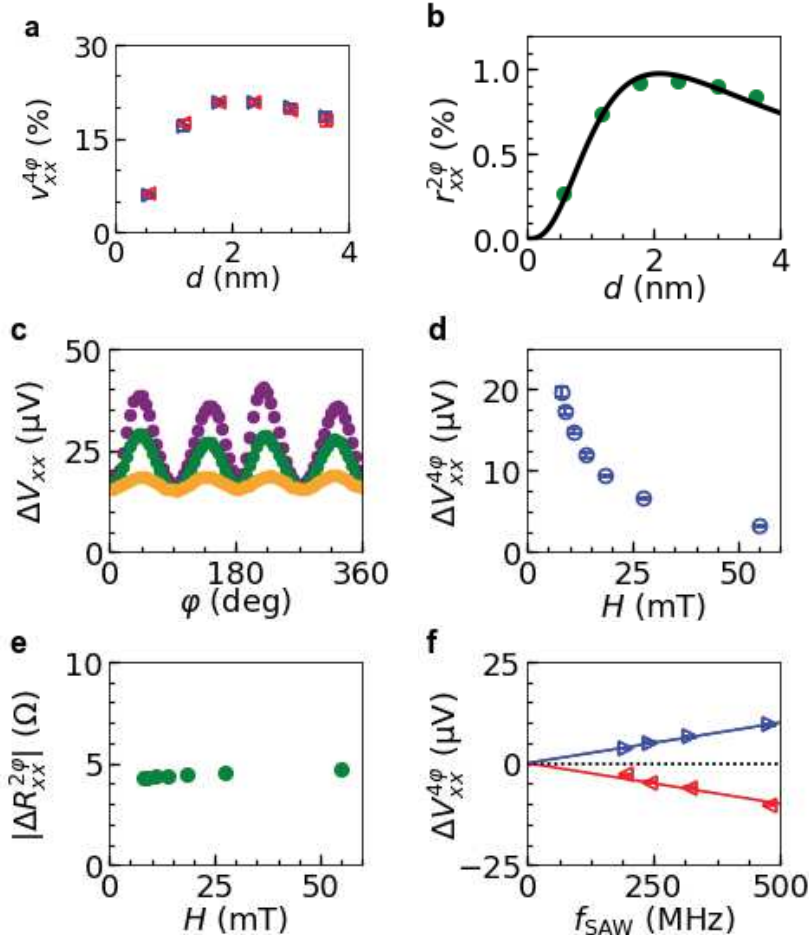}
	\caption{\textbf{X layer thickness, magnetic field and resonance frequency dependence of the acoustic voltage.} (a) Normalized acoustic voltage $v_{xx}^{4\varphi} = \Delta V_{xx}^{4 \varphi} / \Delta V_{xx}^{0}$ plotted against W layer thickness ($d$) for W/CoFeB bilayers. The rf frequency ($f$) and power ($P$) are set to $\sim f_\mathrm{SAW}$ and $\sim10$ dBm, respectively. (b) $d$-dependence of $r_{xx}^{2\varphi}$ of the same system shown in (a). The black line is a fit to the data with Eq.~(\ref{eq:smr}). (c) The field angle ($\varphi_H$) dependence of the acoustic voltage $\Delta V_{xx}$ obtained using various field magnitude ($H$). Purple, green and orange lines are for $H\sim$ 8.0 mT, 14 mT, and 55 mT, respectively. (d) $\Delta V_{xx}^{4\varphi}$ plotted as a function of $H$. (c,d) Data are obtained using rf signal of $f\sim f_\mathrm{SAW}$ and $P\sim$10 dBm applied to IDT1. 
	(e) $H$ dependence of $|\Delta R_{xx}^{2\varphi}|$. 
	(f) The SAW resonance frequency ($f_\mathrm{SAW}$) dependence of $\Delta V_{xx}^{4\varphi}$. The rf power ($P$) is fixed to 10 dBm. 
	The solid lines show linear fits passing through the origin. (c-f) Data presented are obtained using W(2.4)/CoFeB bilayer. 
	The blue (red) triangles in (a,f) represent results when the rf signal is applied to IDT1 (IDT2). The error bars in (a,d,f) show fitting errors of $\Delta V_{xx}$ with Eq.~(\ref{eq:sawphe}). 
	\label{fig:hdep}}
\end{figure}

To identify the origin of $\Delta V_{xx}^{4\varphi}$, we have studied its dependence on the X layer thickness ($d$).
Hereafter, we use $\Delta V_{xx}^{0}$ and $\Delta V_{xx}^{4\varphi}$ to represent the corresponding value at $f \sim f_\mathrm{SAW}$.
As the transmittance of the SAW slightly varies from device to device due to subtle differences in the IDTs,
we normalize $\Delta V_{xx}^{4\varphi}$ with $\Delta V_{xx}^{0}$ and define $v_{xx}^{4\varphi} \equiv \Delta V_{xx}^{4\varphi} / \Delta V_{xx}^{0}$. 
Figure~\ref{fig:hdep}(a) shows the $d$-dependence of $v_{xx}^{4\varphi}$ for W/CoFeB bilayers. 
We find $v_{xx}^{4\varphi}$ takes a maximum at $d \sim 2$ nm.
Interestingly, such $d$-dependence of $v_{xx}^{4\varphi}$ resembles that of the spin Hall magnetoresistance (SMR)\cite{nakayama2013prl,chen2013prb}.
The $d$-dependence of the SMR ratio, $r_{xx}^{2\varphi} \equiv |\Delta R_{xx}^{2\varphi} / R_{xx}^{0} |$ is plotted in Fig.~\ref{fig:hdep}(b).
$\Delta R_{xx}^{2\varphi}$ represents the resistance change when the magnetization of the CoFeB layer is rotated in the $xy$ plane\cite{kim2016prl} and $R_{xx}^{0}$ is the base resistance that does not vary with $\varphi_H$.
Clearly, the $d$-dependence of $v_{xx}^{4\varphi}$ and $r_{xx}^{2\varphi}$ are similar.
According to the drift-diffusion model of spin transport in non-magnetic metal (NM)/ferromagnetic metal (FM) bilayers\cite{chen2013prb,kim2016prl}, the maximum of $r_{xx}^{2\varphi}$ is proportional to the square of the NM layer spin Hall angle ($\theta_\mathrm{SH}$), and the NM layer thickness at the maximum is close to its spin diffusion length ($\lambda_\mathrm{N}$). 
Using the model (see Methods), we obtain $\theta_\mathrm{SH} \sim 0.23$ and $\lambda_\mathrm{N} \sim 0.73$ nm from the $d$-dependence of $r_{xx}^{2\varphi}$ for W, which are in good agreement with previous studies\cite{kim2016prl}.


The similarity in the $d$-dependence of $v_{xx}^{4\varphi}$ and $r_{xx}^{2\varphi}$ suggests that a spin current is generated in the X layer.
The fact that $v_{xx}^{4\varphi}$ is almost absent for Cu/CoFeB bilayers (see Fig.~\ref{fig:sawvol}(d)) further supports this notion: the spin Hall angle of Cu is significantly smaller than that of Pt and W. 
Note, however, that there are a few differences between the acoustic voltage and the SMR.
First, the field-angle dependence of the two is different.
Typically the resistance due to SMR varies as $\cos 2 \varphi_H$ (see for example, Ref.~\cite{nakayama2013prl}), whereas the dominant contribution to the acoustic voltage $\Delta V_{xx}$ varies as $\sin^2 2 \varphi_H$.
Second, $v_{xx}^{4\varphi}$ is more than one order of magnitude larger than $r_{xx}^{2\varphi}$.
Third, we find a striking difference in the magnetic field magnitude ($H$) dependence between the two.
In Fig.~\ref{fig:hdep}(c), we show the $H$ dependence of $\Delta V_{xx}$ vs. $\varphi_H$ for W/CoFeB bilayer. 
As evident, the offset voltage ($\Delta V_{xx}^0$) hardly changes with $H$.
In contrast, the magnitude of $\Delta V_{xx}^{4\varphi}$ increases with decreasing $H$. 
The $H$ dependence of $\Delta V_{xx}^{4\varphi}$, plotted in Figs.~\ref{fig:hdep}(d), shows that $\Delta V_{xx}^{4\varphi}$ scales with $1/H$. 
As a reference, we show in Fig.~\ref{fig:hdep}(e) the $H$ dependence of $|\Delta R_{xx}^{2\varphi}|$.
Contrary to $\Delta V_{xx}^{4\varphi}$, $|\Delta R_{xx}^{2\varphi}|$ is nearly constant against $H$.

To account for these results, we modify the drift-diffusion model of spin transport that is used to describe SMR\cite{chen2013prb}.
First, we include SAW-induced straining of the FM layer and magnetoelastic coupling\cite{gowtham2016prb,lau2017apl}, which cause changes in the magnetization direction with respect to the magnetic field\cite{weiler2011prl,dreher2012prb}. 
Consequently, $\Delta V_{xx}$
acquires an extra factor of $\frac{1}{H} \sin 2 \varphi_H$ compared to the resistance change that originates from SMR.
(See supplementary material section~\ref{sec:supp:results} where we show that $\Delta V_{xx}^{4 \varphi}$ is absent for W/NiFe bilayer due to the small magnetoelastic coupling of NiFe.)
Next, to generate a (rectified) dc current, the spin current must vary in time and space such that it couples to the motion of magnetic moments driven by the SAW-induced strain.
We find the following form of spin current $j_{\mathrm{s}, z}^y$ (electron spin orientation along $y$ and flow along $z$) produces a rectified dc current and accounts for the experimental results:
\begin{equation}
\begin{aligned}
j_{\mathrm{s}, z}^y = A \frac{\partial u_x}{\partial t},
\label{eq:phononspin1}
\end{aligned}
\end{equation}
where $u_{x}$ is the lattice displacement along the wave propagation direction ($x$).
$A$ is a prefactor that determines the spin current generation efficiency and is proportional to $\lambda_\mathrm{so}$, the SOI. 

The spin current $j_{\mathrm{s}, z}^y$ generated in the NM layer drifts to the NM/FM interface and causes spin accumulation.
The accumulated spin at the interface causes a back flow of spin current within the NM layer, which is converted to electrical current via the inverse spin Hall effect\cite{saitoh2006apl}. 
The amount of spin accumulation at the interface depends on the direction of the FM layer magnetization due to the action of spin transfer torque\cite{nakayama2013prl,chen2013prb}, thus causing the $\varphi_H$ dependent acoustic voltage.
The resulting acoustic voltage reads (see supplementary material section~\ref{sec:supp:model})
\begin{equation}
\begin{aligned}
\Delta V_{xx}  \approx c \lambda_{\mathrm{so}}^2 K(d) f_\mathrm{SAW} P \ \mathrm{sgn}(k) \frac{b}{H M_\mathrm{S}} \sin^2 2 \varphi_H,
\label{eq:jcxx}
\end{aligned}
\end{equation}
where $c$ is a constant that depends on the material and the geometry of the device, $K(d)$ characterizes the $d$-dependence similar to that of the SMR (see Eq.~(\ref{eq:smr})), $k$ is the wave vector of the Rayleigh-type SAW (sgn$(x)$ takes the sign of $x$), and $b$ and $M_\mathrm{S}$ are, respectively, the magnetoelastic coupling constant and the saturation magnetization of the FM layer.

Equation~(\ref{eq:jcxx}) captures many features of the acoustic voltage found in the experiments.
As evident, $\Delta V_{xx}$ varies as $\sin^2 2\varphi_H$. 
The coefficient of $\sin^2 2\varphi_H$ in Eq.~(\ref{eq:jcxx}), equivalent to $\Delta V_{xx}^{4\varphi}$, changes its sign upon reversal of the wave propagation direction (defined by the sign of $k$), scales with $\frac{1}{H}$ and $P$, and is proportional to the square of the spin orbit coupling of the NM layer, and thus independent of the sign of the NM layer spin Hall angle.
The thickness dependence of $\Delta V_{xx}^{4\varphi}$, coded in $K(d)$, is in relatively good agreement with the experimental results.
We have also studied the $f_\mathrm{SAW}$ dependence of $\Delta V_{xx}^{4\varphi}$ for W/CoFeB bilayer; the results are plotted in Fig.~\ref{fig:hdep}(f).
As evident, $\Delta V_{xx}^{4\varphi}$ scales with $f_\mathrm{SAW}$. 
We emphasize that Eq.~(\ref{eq:phononspin1}) is the only form of spin current that can account for these results.
Note that the linear dependence of $\Delta V_{xx}^{4\varphi}$ with $f_\mathrm{SAW}$ excludes contributions from spin-dependent inertial force\cite{matsuo2017jpsj} and related effects in the presence of SOI\cite{funato2018jpsj}, which are proportional to higher order of $f_\mathrm{SAW}$.

\begin{figure}[b]
	\includegraphics[scale=1.0]{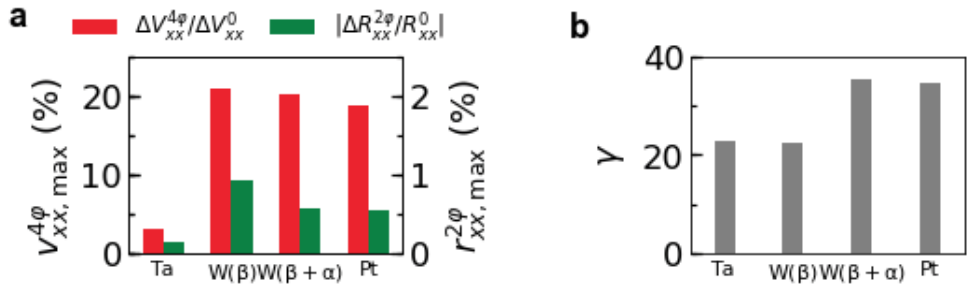}
	\caption{\textbf{Efficiency to generate lattice motion induced spin current.} (a,b) Maximum values of the normalized acoustic voltage $v_{xx, \mathrm{max}}^{4\varphi} \equiv \big|\Delta V_{xx}^{4 \varphi} / \Delta V_{xx}^{0} \big\vert_\mathrm{max}$ (red bars) and the maximum SMR ratio $r_{xx, \mathrm{max}}^{2\varphi} \equiv \big|\Delta R_{xx}^{2\varphi} / R_{xx}^{0} \big|_\mathrm{max}$ (green bars) (a) and their ratio $\gamma \equiv \frac{v_{xx, \mathrm{max}}^{4\varphi}}{r_{xx, \mathrm{max}}^{2\varphi}}$ (b) obtained for X/CoFeB (X=Ta, W, and Pt) bilayers and CoFeB/W bilayers. W($\beta$) and W($\beta + \alpha$) represent W/CoFeB and CoFeB/W bilayers, respectively. 
	\label{{fig:matdep}}}
\end{figure}

These results therefore demonstrate that the lattice motion induces a spin current.
Recent studies have shown that spin-rotation coupling\cite{matsuo2013prb,takahashi2016nphys} can induce spin accumulation in the NM layer, which results in generation of spin current if the NM layer thickness is larger than the SAW decay length (typically, of the order the SAW wavelength, which is a few $\upmu$m here)\cite{kobayashi2017prl}.
To clarify the role of spin-rotation coupling, we have studied $\Delta V_{xx}^{4\varphi}$ of inverted structures, CoFeB/W bilayers.
In both W/CoFeB and CoFeB/W bilayers, spin-rotation coupling induces spin density in the W layer, which can cause a flow of spin current toward the CoFeB layer as the latter can act as a spin sink.
If such spin current were to flow, the flow direction will be opposite for the normal (W/CoFeB) and inverted (CoFeB/W) structures and consequently results in $\Delta V_{xx}^{4\varphi}$ with opposite sign.
We find that the signs of $\Delta V_{xx}^{4\varphi}$ for W/CoFeB and CoFeB/W bilayers are the same, demonstrating  that spin-rotation coupling is not the source of spin current (see supplementary material sections~\ref{sec:supp:results} and \ref{sec:supp:artifact}).
For the same reason, we can rule out SAW-induced spin pumping\cite{weiler2011prl,uchida2011apl} from the CoFeB layer and the inverse spin Hall effect of the W layer.
This is also supported by the fact that the signs of $\Delta V_{xx}^{4\varphi}$ for W/CoFeB and Pt/CoFeB bilayers are the same (see Fig.~\ref{fig:sawvol}) albeit the difference in the sign of $\theta_\mathrm{SH}$ for W and Pt.

In Fig.~\ref{{fig:matdep}}(a), we summarize the maximum value of $v_{xx}^{4\varphi}$ and $r_{xx}^{2\varphi}$  when $d$ is varied, denoted as $v_{xx, \mathrm{max}}^{4\varphi}$ and $r_{xx, \mathrm{max}}^{2\varphi}$, respectively, for each bilayer (X=Ta, W, Pt).
Results from the CoFeB/W bilayers are included.
Note that the structure of W depends on the growth condition: from the film resistivity\cite{petroff1973jap,pai2012apl}, we consider W forms a highly-resistive $\beta$-phase in W/CoFeB bilayer whereas it is a mixture of the $\beta$-phase and the low-resistivity crystalline $\alpha$-phase in CoFeB/W bilayer.
Consequently, the SMR ratio ($r_{xx, \mathrm{max}}^{2\varphi}$) is smaller for the latter due to the smaller $\theta_\mathrm{SH}$\cite{pai2012apl,liu2015apl,sui2017prb}.
Interestingly, we find that $v_{xx, \mathrm{max}}^{4\varphi}$ takes nearly the same value for the two bilayers, indicating that there are factors other than $\theta_\mathrm{SH}$ that sets the magnitude of $v_{xx, \mathrm{max}}^{4\varphi}$.
In Fig.~\ref{{fig:matdep}}(b), we plot the ratio $\gamma \equiv \frac{v_{xx, \mathrm{max}}^{4\varphi}}{r_{xx, \mathrm{max}}^{2\varphi}}$ to characterize such contribution. 
We find $\gamma$ is significantly larger for bilayers with Pt and ($\beta$+$\alpha$)-W (CoFeB/W) than that with $\beta$-W (W/CoFeB) and Ta.
Since the former two layers are textured whereas the latter two are highly disordered (\textit{i.e.}, amorphous-like), we consider the texture of the films may influence $\gamma$.
Little correlation is found between $\gamma$ and the bulk modulus of the X layer.

Finally, we discuss the source of spin current that scales with the time derivative of lattice displacement (Eq.~(\ref{eq:phononspin1})).
First, a conventional mechanism would be to consider internal electric field associated with the SAW and the resulting spin Hall effect of the NM layer.
There are two major sources of internal electric field.
One is the piezoelectric field ($E_\mathrm{p}$) localized at the film/substrate interface.
Spin current generated from $E_\mathrm{p}$ can only reach the NM/FM interface when the film thickness is smaller than $\lambda_\mathrm{N}$.
The thickness dependence of $v_{xx}^{4\varphi}$ (Fig.~\ref{fig:hdep}(a)) rules out such contribution.
The other source is the time varying electric field ($E_\mathrm{b}$) caused by the motion of ions\cite{pippard1955philmag,holstein1959pr,blount1959pr}.
 $E_\mathrm{b}$ is uniform along the film normal as long as the film thickness is significantly smaller than the SAW decay length.
In general, $E_\mathrm{b}$ is screened by the conduction electrons in metallic films: we infer it generates negligible spin current. 
With the current understanding, we consider it is difficult to quantitatively account for the experimental results with the combination of the SAW induced electric field and the spin Hall effect.
Second, Eq.~(\ref{eq:phononspin1}) can be derived assuming the following interaction\cite{pavlov1966sovphys,romano2008prb}: $H_\mathrm{int} = s \bm{u} \cdot (\bm{p} \times \bm{\sigma})$, where $s$ is a constant, $\bm{u}$ is the lattice displacement vector, and $\bm{p}$ and $\bm{\sigma}$ are electron momentum and spin orientation, respectively. 
This interaction derives from the SOI\cite{pavlov1966sovphys,romano2008prb} and the coefficient $s$ is proportional to $\lambda_\mathrm{so}$, similar to the relation between $\theta_\mathrm{SH}$ and $\lambda_\mathrm{so}$.
$H_\mathrm{int}$ resembles the Rashba Hamiltonian\cite{bychkov1984jetp} but can exist here since the inversion symmetry is broken by the dynamical lattice displacement ${\bm u}$.
Further studies are required, however, to justify the presence of such Hamiltonian.
Third, the time derivative of the lattice displacement can cause changes in the Berry curvature of electron wave function.
Indeed, theoretical studies have identified the right hand side of Eq.~(\ref{eq:phononspin1}) as the Berry electric field\cite{sundaram1999prb,chaudhary2018prb}. 
It remains to be seen whether spin current emerges from the Berry electric field under strong SOI.
Finally, the phonon angular momentum\cite{lfzhang2014prl,garanin2015prb,holanda2018nphys} may contribute to the generation of spin current.
Similar to the spin Seebeck effect\cite{uchida2010nmat}, where the spin angular momentum of magnons are transferred to electrons, the angular momentum of phonons (i.e. sound waves) can be transferred to the electrons and induce spin current.
The efficiency of such process must be addressed to assess its contribution.

In summary, we have shown that spin current is directly created from lattice motion associated with surface acoustic wave (SAW).
Such acoustic spin Hall effect is observed in non-magnetic metal (NM)/ferromagnetic metal (FM) bilayers through a field-dependent dc acoustic voltage. 
The acoustic voltage roughly scales with the square of the spin Hall angle of the NM layer and is proportional to the SAW frequency.
The NM layer thickness dependence of the acoustic voltage is similar to that of the spin Hall magnetoresistance.
Using a diffusive spin transport model, we show that such characteristics of the acoustic voltage can be accounted for when a spin current that scales with the time derivative of lattice displacement is generated in the NM layer.
Possible sources of such spin current include a Berry electric field associated with time varying Berry curvature and/or an unconventional SOI-mediated spin-lattice interaction that resembles the form of Rashba Hamiltonian.
The efficiency to generate spin current, represented by the maximum acoustic voltage, also seems to depend on a factor related to the film texture; the efficiency is nearly the same for amorphous-like $\beta$-W and textured Pt despite the difference in their spin Hall angle.
The finding of the acoustic spin Hall effect thus implies a mechanism that facilitates an SOI mediated coupling of electron spins and a rotating lattice.
Further studies are required to unveil the microscopic mechanism to describe such coupling.

\section{Materials and Methods}
\subsection{\label{sec:methods:device}Sample preparation}
Radio frequency magnetron sputtering is used to deposit the films on piezoelectric Y+128$^\circ$-cut LiNbO$_3$ substrates. 
The film structure is sub./X($d$)/CoFeB(1)/MgO(2)/Ta(1) with X=W, Pt, Ta and Cu (thickness in unit of nanometers).
The inverted structure is sub./MgO(2)/CoFeB(1)/X($d$)/MgO(2)/Ta(1) with X=W.
The MgO(2)/Ta(1) layers serve as a capping layer to prevent oxidation of the films.
For bilayers with X=Pt and Cu, a 0.5 nm thick Ta layer is inserted before deposition of X to promote their smooth growth. 
Hall bars are formed from the films using optical lithography and Ar ion etching. 
Subsequently, we use optical lithography and a liftoff process to form interdigital transducers (IDTs) and electrodes made of Ta(5)/Cu(100)/Pt(5). 

\begin{figure}[h]
	\includegraphics[scale=1.0]{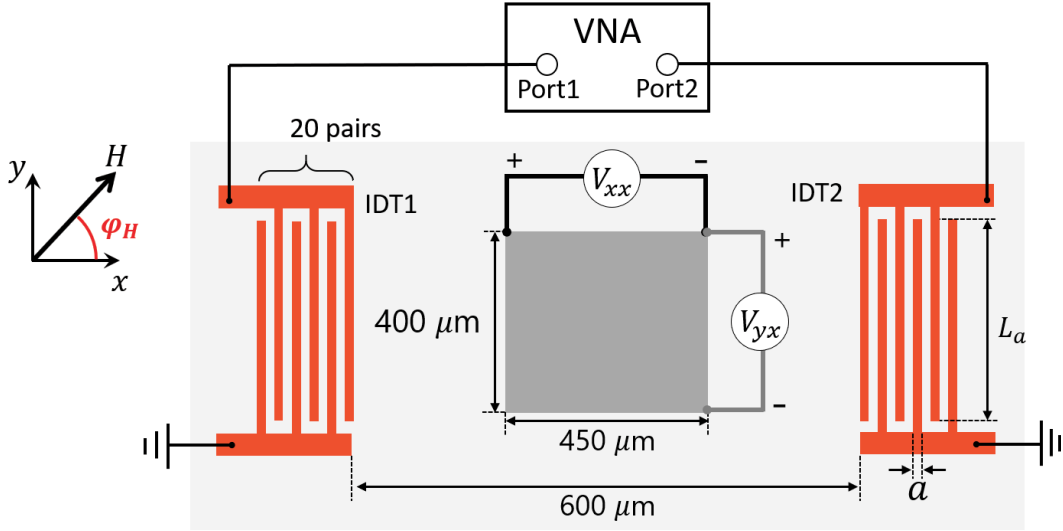}
	\caption{Schematic illustration of the SAW device. The orange structure represent the IDTs and the dark gray area show the film. 
	\label{fig:SAWdevice}}
\end{figure}

Schematic illustration of the SAW device and definition of the coordinate system are shown in Fig.~\ref{fig:SAWdevice}. The distance of the two IDTs is $\sim$ 600 $\upmu$m and each IDT has 20 pairs of single-type fingers. The width and gap of the fingers are set to $a$: the corresponding SAW wavelength is $\sim 4a$.
The finger overlap, \textit{i.e.}, the SAW aperture ($L_\mathrm{a}$), is fixed to $\sim 450$ $\upmu$m. 
A Hall bar made of the film is placed at the center of the two IDTs. 
The length and width of the Hall bar are set to $\sim 450$ $\upmu$m and $\sim 400$ $\upmu$m, respectively. 

We vary $a$ to change the SAW resonance frequency ($f_\mathrm{SAW}$).
$a$ is fixed to $\sim5$ $\upmu$m for most of the results shown, which gives $f_\mathrm{SAW} \sim 194$ MHz.
In Fig.~\ref{fig:hdep}(f), we vary $a$ to change $f_\mathrm{SAW}$: $a$ is set to $\sim$5, $\sim$4, $\sim$3, $\sim$2 $\upmu$m to obtain $f_\mathrm{SAW}$ of $\sim$194, $\sim$242, $\sim$321, $\sim$479 MHz, respectively.

\subsection{\label{sec:methods:volt}Voltage measurements}
The longitudinal (along $x$) and transverse (along $y$) voltages, defined as $V_{xx}$ and $V_{yx}$, respectively, are measured during the SAW excitation.
To extract the voltage originating from the SAW, we subtract the average voltage measured under off-resonance conditions, defined as $\langle {V}_{xx(yx)}^\mathrm{off} \rangle$.
$\langle {V}_{xx(yx)}^\mathrm{off} \rangle$ is obtained as follows.
Under a fixed magnetic field and rf power, we study the frequency ($f$) dependence of $V_{xx(yx)}$.
$V_{xx(yx)}$ takes a peak when $f \sim f_\mathrm{SAW}$. 
We choose frequencies ($f_\mathrm{off}$) that are outside the peak structure of $V_{xx(yx)}$, typically a few tens of MHz away from $f_\mathrm{SAW}$ (see Fig.~\ref{fig:schematics}(c) for a typical transmission spectra).
$\langle {V}_{xx(yx)}^\mathrm{off} \rangle$ is the average value of $V_{xx(yx)}$ measured at several $f_\mathrm{off}$.
$\langle {V}_{xx(yx)}^\mathrm{off} \rangle$ is subtracted from the measured voltage $V_{xx(yx)}$ at frequency $f$ to obtain the acoustic voltage $\Delta V_{xx(yx)} \equiv V_{xx(yx)} - \langle {V}_{xx(yx)}^\mathrm{off} \rangle$.
$\langle {V}_{xx(yx)}^\mathrm{off} \rangle$ is always measured prior to the measurement of $V_{xx(yx)}$ at frequency $f$.
Voltage measurements at each condition are repeated 5-100 times to improve the signal to noise ratio.

\subsection{\label{sec:methods:smr}Spin Hall magnetoresistance}
In the main text, we have used $\Delta R_{xx}^{2\varphi}$, the resistance change when the magnetization of the CoFeB layer is rotated in the $xy$ plane, to estimate SMR.
$\Delta R_{xx}^{2\varphi}$ is equal to the sum of the SMR and the anisotropic magnetoresistance (AMR). 
Since the latter is significantly smaller than the former for the system under study\cite{kim2016prl}, we assume $\Delta R_{xx}^{2\varphi}$ represents the SMR.
To obtain the SMR more accurately, it is customary to measure the resistance change when the magnetization of the CoFeB layer is rotated in the $yz$ plane\cite{nakayama2013prl}, defined as $\Delta R_{xx}^\mathrm{smr}$.
We have verified that $\Delta R_{xx}^{2\varphi}$ and $\Delta R_{xx}^\mathrm{smr}$ take similar value, justifying the assumption that $\Delta R_{xx}^{2\varphi} / R_{xx}^0$ represents the SMR. 

The X layer thickness dependence of the spin Hall magnetoresistance is fitted using the following equation\cite{nakayama2013prl,chen2013prb}:
\begin{equation}
\begin{aligned}
\frac{\Delta R_{xx}^{2 \varphi}}{R_{xx}^0} &= \frac{\theta_\mathrm{SH}^2}{1+\zeta} K(d),\\
&K(d) \equiv \frac{\lambda_\mathrm{N}}{d} \tanh^2 \frac{d}{2 \lambda_\mathrm{N}} \tanh \frac{d}{\lambda_\mathrm{N}},
\label{eq:smr}
\end{aligned}
\end{equation}
where $\zeta \equiv \frac{\rho_\mathrm{N} t_\mathrm{F}}{\rho_\mathrm{F} d}$, $\rho_\mathrm{F}$ and $t_\mathrm{F}$ are the resistivity and thickness of the FM (=CoFeB) layer, respectively and $\rho_\mathrm{N}$ is the resistivity of the X layer. Here we have assumed a transparent X/FM interface for spin transmission and neglected the effect of longitudinal spin absorption of the FM layer\cite{kim2016prl}.  
The base longitudinal resistance $R_{xx}^0$ is defined as the resistance when the magnetization of the FM layer points along the $y$-axis. 
For fitting the data (Fig.~\ref{fig:hdep}(b)) with Eq.~(\ref{eq:smr}), we have used $\rho_\mathrm{N} \sim 147$ $\upmu\Omega\cdot$cm and $\rho_\mathrm{F} \sim 160$ $\upmu\Omega\cdot$cm.

\begin{acknowledgments}
This work was partly supported by JSPS Grant-in-Aid for Specially Promoted Research (15H05702), and the Center of Spintronics Research Network of Japan.
\end{acknowledgments}


\bibliography{reference_061220}

\end{document}